\documentstyle[11pt]{article}
\begin{document}
 \newcommand{\be}{\begin{equation}} \newcommand{\fe}{\end{equation}}
\newcommand{\eqn}{\label}\newcommand{\bel}{\begin{equation}\label}

 
\def\spose#1{\hbox to 0pt{#1\hss}}\def\lta{\mathrel{\spose{\lower 3pt\hbox
{$\mathchar"218$}}\raise 2.0pt\hbox{$\mathchar"13C$}}}  \def\gta{\mathrel
{\spose{\lower 3pt\hbox{$\mathchar"218$}}\raise 2.0pt\hbox{$\mathchar"13E$}}}

\noindent

{\bf OLD AND NEW PROCESSES OF VORTON FORMATION}

\medskip
{\bf Brandon Carter}

\medskip
{\bf D.A.R.C., Observatoire de Paris}

{\bf 92 Meudon, France}

\medskip
{ Contribution to 35th Karpacs meeting  (ed. J. Kowalski-Glikman)

Polanica, Poland, February 1999.}

\bigskip
{\bf Astract} {\it After a brief explanation of the concept of a vorton,
quantitative estimates of the vorton population that would be produced
in various cosmic string scenarios are reviewed. Attention is drawn
to previously unconsidered mechanisms that might give rise to much
more prolific vorton formation that has been envisaged hitherto.}

\bigskip \parindent =2 cm

This review is an updated version of a previous very brief 
overview\cite{overview} of the theory of vortons, meaning equilibrium 
states of cosmic string loops, and of the cosmological processes by 
which they can be produced in various scenarios. The main innovation
here is to draw attention to the possibility of greatly enhanced
vorton formation in cases for which the cosmic string current is of the
strictly chiral type~\cite{CP99} that arises naturally in certain
kinds of supersymmetric field theory.

It is rather generally accepted\cite{book} that among the conceivable
varieties of topological defects of the vacuum that might have been generated
at early phase transitions, the {\it vortex type} defects describable on a
macrosopic scale as {\it cosmic strings} are the kind that is most likely to
actually occur -- at least in the post inflationary epoch -- because the other
main categories, namely monopoles and walls, would produce a catastrophic
cosmological mass excess. Even a single wall stretching accross a Hubble
radius would by itself be too much, while in the case of monopoles it is their
collective density that would be too high unless the relevant phase transition
occurred at an energy far below that of the G.U.T. level, a possibility that
is commonly neglected on the grounds that no monopole formation occurs in the
usual models for the transitions in the relevant range, of which the most
important is that of electroweak symmetry breaking. 

The case of cosmic strings is different. One reason is that -- although they
are not produced in the standard electroweak model -- strings are indeed
produced at the electroweak level in many of the commonly considered (e.g.
supersymmetric) alternative models. A more commonly quoted reason why the case
of strings is different, even if they were formed at the G.U.T level, is that
-- while it may have an important effect in the short run as a seed for galaxy
formation -- such a string cannot be cosmologically dangerous just by itself,
while a distribution of cosmic strings is also cosmologically harmless because
(unlike ``local'' as opposed to ``global'' monopoles) they will ultimately
radiate away all their energy and disappear. However while this latter
consideration is indeed valid in the case of ordinary Goto-Nambu type strings,
it was  pointed out by Davis and Shellard\cite{DS} that it need not apply to
``superconducting'' current-carrying strings of the kind originally introduced
by Witten\cite{witten}. This is because the occurrence of stable currents
allows loops of string to be stabilized in states known as ``vortons'', so
that they cease to radiate. 

The way this happens  is that the current, whether timelike or spacelike,
breaks the Lorentz invariance along the string worldsheet
\cite{for,mal,neutral,enon0}, thereby leading to the possibility of rotation,
with velocity $v$ say. The centrifugal effect of this rotation, may then
compensate the string tension $T$ in such a way as to produce an equilibrium
configuration, i.e. what is known as a {\it vorton}, in which 
\be T= v^2 U \, , \eqn{1}\fe
where $U$ is the energy per unit length in the corotating rest 
frame\cite{ring,CPG96}. Such a vorton state will be stable, at least
classically, if it minimises the energy for given values of the pair of
conserved quantities characterising the current in the loop, namely the phase
winding number $N$ say, and the corresponding particle number $Z$ say, whose
product determines the mass $M$ of the ensuing vorton state according to a
rough order of magnitude formula of the form
\be M\approx\vert NZ\vert^{1/2} m_{\rm x} \eqn{2}\fe
where $m_{\rm x}$ is the relevant Kibble mass, whose square is the zero current
limit value of both $T$ and $U$. If the current is electromagnetically
coupled, with charge coupling constant $e$, then there will be a 
corresponding vorton charge $Q=Ze$.

Whereas the collective energy density of a distribution of non-conducting
cosmic strings will decay in a similar manner to that of a radiation gas, in
contrast for a  distribution of relic vortons the energy density will scale
like that of ordinary matter. Thus, depending on when and how efficiently 
they were formed, and on how stable they are in the long run, such vortons 
might eventually come to dominate the density of the universe.  It has been
rigorously established\cite{stabgen,stab,stabwit} that circular vorton
configurations of this kind will commonly (though not always) be stable in 
the dynamic sense at the classical level, but very little is known so far 
about non-circular configurations or about the question of stability against 
quantum tunnelling effects, one of the difficulties being that the latter is 
likely to be sensitively model dependent.

In the earliest crude quantitative estimates\cite{DS,vorton} of the likely
properties of a cosmological vorton distribution produced in this way, it 
was assumed not only that the Witten current was stable against leakage by
tunnelling, but also that the mass scale $m_\sigma$ characterising the
relevant carrier field was of the same order of magnitude as the Kibble mass
scale $m_x$ characterising the string itself, which will normally be given
approximately by the mass of the Higgs field responsible for the relevant
vacuum symmetry breaking. The most significant development in the more
detailed investigations carried out more recently\cite{Moriand95,BCDM96} was
the extension to cases in which $m_\sigma$ is considerably smaller than 
$m_x$.  A rather extreme example that immediately comes to mind is that for 
which $m_x$ is postulated to be at the G.U.T. level, while $m_\sigma$ is at 
the electroweak level in which case it was found that the resulting vorton 
density would be far too low to be cosmologically significant.

The simplest scenarios are those for which (unlike the example just quoted)
the relation 
\be \sqrt{m_\sigma}\gta m_{\rm x} \eqn{3}\fe
is satisfied in dimensionless Planck units as a rough order of magnitude
inequality. In this case the current condensation would have ocurred during
the regime in which (as pointed out by Kibble\cite{kibble} in the early
years of cosmic string theory) the dynamics was dominated by friction damping.
Under these circumstances, acording to the standard picture\cite{book},
the string distribution will consist of wiggles and loops of which
the most numerous will be the shortest, characterised by a length scale
$\xi$ say below which smaller scale structure will have been smoothed
out by friction damping. The number density $n$ of these smallest and
most numerous loops will be given by the (dimensionally obvious)
formula 
\be n\approx\xi^{-3}\, , \eqn{3a}\fe
in which the smoothing length scale $\xi$ itself is given by 
\be \xi\approx \sqrt{t\tau}\, ,\eqn{3b}\fe
where $\tau$ is the relevant friction damping timescale and $t$ 
is the cosmological time, which, using Planck units, will be
expressible in terms of the cosmological temperature $\Theta$ by 
\be t\approx\Theta^{-2} \, ,\eqn{3c}\fe
in the  radiation dominated epoch under consideration.
According to the usual description of the friction dominated 
epoch~\cite{VV85,book}, the relevant damping timescale will be given by
\be \tau\approx m_{\rm x}^{\, 2}\Theta^{-3}\, ,\eqn{3d}\fe
from which it can be seen that the smoothing lengthscale $\xi$ that
characterises the smallest and most numerous string loops will
be given roughly by the well known formula
\be \xi\approx m_{\rm x}\Theta^{-5/2}\, .\eqn{3e}\fe

At the time of condensation of the current carrier
field on the strings, when the temperature reaches a value 
$\Theta\approx m_\sigma$, the corresponding thermal fluctuation
wavelength $\lambda$ will be given by 
\be \lambda\approx m_\sigma^{-1}\, .\eqn{3f}\fe 
Taken around the circumference, of order $\xi$, of a typical
small string loop, the number of such fluctuation wavengths will
be of order $\xi/\lambda$. In the cases considered
previously~\cite{Moriand95,BCDM96}
it was assumed that the fluctuations would
be randomly orientated and would therefore tend to cancel each other
out so that, by the usual kind of random walk process the net
particle and winding numbers taken around the loop as a whole would
be expected to be of the order of the square root of this number
of wavelengths, i.e. one would typically obtain
\be N \approx Z \approx \sqrt{\xi/\lambda}\, .\eqn{4a}\fe
However a new point to which I would like to draw attention here is
that the random walk cancellation effect will not apply in case for 
which the current is of strictly chiral type
so that the string dynamics is of the kind whose special integrability
properties have recently been pointed out~\cite{CP99}.
This case arises~\cite{witten} when the string current is attributable
to (necessarily uncharged) fermionic zero modes moving in an
exlusively rightwards (or exclusively leftwards) direction.
In such a case a loop possibility of cancellation between left moving
and right moving fluctuations does not arise so that (as in the ordinary
kind of diode rectifier circuit used for converting alternating current to
direct curent) there is an effective filter ensuring that the fluctuations
induced on the string will all have the same orientation. In such a case
only one of the quantum numbers in the formula (\ref{2}) will be
independent, i.e. they will be restricted by a relation of the form
$N=Z$, and their expected value will be of the order of the total
number of fluctuation wavelengths round the loop (not just the square
root thereof as in the random walk case). In such a strictly chiral case
the formula (\ref{2}) should therefore be evaluated using an estimate
of the form
\be N=Z\approx \xi/\lambda\, ,\eqn{4b}\fe
instead of (\ref{4a})

Whereas even smaller loops will have been entirely destroyed by
the friction damping process, those that are present at the time
of the current condensation can survive as vortons, whose number
density will be reduced in inverse proportion to the comoving
volume, i.e. proportionally to $\Theta^3$, relative to the initial 
number density value given by (\ref{3a}) when $\Theta\approx 
m_{\sigma}$.
Thus (assuming the current on each string is strictly conserved
during the subsequent evolution)
when the cosmological temperature has fallen to a 
lower value $\Theta \ll m_{\sigma}$, the
expected number density $n$ of the vortons will be
given  as a constant fraction of
the corresponding number density  $\approx \Theta^3$ of black body photons by
the rough order of magnitude formula 
\be {n\over\Theta^3}\approx \Big( {\sqrt{m_\sigma}\over m_x}\Big)^3
 m_\sigma^{\,3} \, .\eqn{6}\fe 

In the previously considered cases~\cite{Moriand95,BCDM96}, 
for which the random walk formula (\ref{4a}) applies, 
the typical value of the quantum numbers of vortons in the resulting
population will be given very roughly by 
\be N\approx Z\approx  m_{\rm x}^{\,1/2} m_\sigma^{-3/4} \, ,\eqn{4} \fe 
which by (\ref{2}) implies a typical vorton mass given by 
\be M\approx\Big({m_x\over\sqrt{ m_\sigma}}\Big)^{3/2} \, ,\eqn{5}\fe
which, in view of (\ref{3}), will never exceed the Planck mass. 
It follows in this case that, in order to avoid producing a cosmological 
mass excess, the value of $m_\sigma$ in this formula should not exceed 
a limit that works out to be of the order of $10^{-9}$, and the limit 
is even be smaller, $m_\sigma\ll 10^{-11}$, when the two scales 
$m_\sigma$ and $m_x$ are comparable.

The new point to which I wish to draw attention here is that for
the strictly chiral case, as characterised by (\ref{4b}) instead of
(\ref{4a}), the formula (\ref{2}) for the vorton mass gives a 
typical value
\be M \approx m_{\rm x}^{\,2} m_{\sigma}^{-3/2}\, ,\eqn{5b}\fe
which is greater than the what is given by the usual formula
(\ref{5}) by a factor $m_{\rm x}^{\,1/2}m_{\sigma}^{-3/4}$.
Although the vorton to photon number density ratio (\ref{6}) will
not be affected, the corresponding mass density $\rho=Mn$ of 
the vorton distribution will be augmented by the same factor 
$m_{\rm x}^{\,1/2}m_{\sigma}^{-3/4}$. This augmentation factor 
will be expressible simply as $m_\sigma^{-1/4}$  when the two scales 
$m_\sigma$ and $m_x$ are comparable, in which case the requirement 
that a cosmological mass should be avoided leads to the rather
severe limit $m_\sigma\lta 10^{-14}$. This mass limit works out to be
of the order of a hundred TeV, which is within the range that
is commonly envisaged for the electroweak symmetry breaking 
transition. 

The foregoing conclusion can be construed as meaning that
if strictly chiral current carrying strings
were formed (within the framework of some generalised, presumably
supersymmetric, version of the Standard electroweak model) during
the electroweak symmetry breaking phase transition, then the
ensuing vorton population might conceivably constitute a significant
fraction of the cosmological dark matter distribution in the
universe. Although, according to (\ref{6}), the number density
of such chiral vortons would be rather low, their typical mass,
as given according to (\ref{5b}) by $M\approx \sqrt{m_\sigma}$
would be rather large, about $10^{-7}$ in Planck units, which
works out as about $10^{9}$ TeV.
 
An alternative kind of scenario that naturally comes to mind
is that in which the cosmic strings themselves were formed at an energy
scale $m_{\rm x}$ in the GUT range (of the order of $10^{-3}$
in Planck units) but in which the current did not condense on the
string until the thermal energy scale had dropped to a 
value $m_\sigma$ that was nearer the electroweak value (below
of $10^{-10}$ in Planck units). However since this very much
lower condensation temperature would be outside the friction
dominated range characterised by (\ref{3}) would not be applicable.
Preliminary evaluations of (relatively inefficient) vorton production 
that would arise from current condensation
after the end of the friction dominated period are already
available~\cite{BCDM96} for the usual random walk case, but analogous
estimates for aumentation that might arise in the strictly chiral case
have not yet been carried out. The reason why it is not so easy to
evaluate the consequences
of current condensation after the end of the friction dominated
epoch (when radiation damping becomes the main dissipation mechanism)
is that most of the loops present at the time of the current condensation
would have been be too small to give vortons stable against quantum
decay processes, a requirement which imposes a lower limit
\be M\gta m_{\rm x}^{\, 2}/m_\sigma
\eqn{ql}\fe
 on the mass of a viable vorton.
This condition is satisfied automatically by the masses estimated
in the manner described above for vortons formed by condensation
during the friction dominated era characterised by (\ref{3}).
On the other hand when (\ref{3}) is not satisfied -- in which case
the lower limit (\ref{ql}) will evidently exceed the Planck mass -- then
the majority of loops present at the time of the carrier condensation
phase transition at  the temperature $\Theta\approx m_\sigma$ will
not acquire the rather large quantum number values that would be
needed to make them ultimately viable as vortons.  It is not at all easy
to obtain firmly conclusive  estimates of the small fraction that will
satisfy this viability condition. However it should not be too difficult
to carry out an adaptation to the strictly chiral case of the kind of
tentative provisional estimates (based on simplifying assumptions whose
confirmation will require much future work) that have already been
provided~\cite{BCDM96} for the generic of currennts built up
by the usual random walk process.

The possibility of strictly chiral current formation is not the only
mechanism whereby vorton formation might conceivably be augmented
relative to what was predicted on the basis~\cite{BCDM96} of the previous
estimates, which took no account of electromagnetic effects. There cannot
be any electromagnetic coupling in the strictly chiral case~\cite{CP99},
and in other cases where electromagnetic coupling will be typically
be present it has been shown~\cite{P93} that it will usually have
only a minor perturbing effect on the vorton equilibrium states.
However it has recently been remarked~\cite{DD98} that even though
the averaged ``direct'' current that is relevant for vorton formation
may be small, the local `alternating' current can have a sufficiently
large amplitude, $I$ say, for its interaction with the surrounding black
body radiation plasma to provide the dominant friction damping
mechanism, with a damping scale that instead of (\ref{3d}) will be
given in rough order of magnitude by
$\tau\approx m_{\rm x}^{\, 2} I^{-1}\Theta^{-2}$.
This means that instead of being restricted to the very early epoch
when cosmological temperature was above Kibble limit value,
i.e. when $\Theta\gta \sqrt{m_{\rm x}}$, the period of friction
domination can be extended indefinitely if the current amplitude
satisfies $I\gta m_{\rm x}^{\, 2}$, a requirement that is easily
compatible with Witten's~\cite{witten} current saturation bound
$I\lta e m_{\rm x}$ that applies (where the $e\simeq 1/\sqrt{137}$
is the electromagnetic charge coupling constant) and in most cases
even with the more severe limit $I\lta e m_\sigma$ that applies
in cases for which instead of arising as a bosonic condensate, the
current is due to femionic zero modes. Such a tendency to prolongation of
friction dominance will presumably delay the decay of small scale
loop structure and so may plausibly be expected to augment the
efficiency of vorton formation in cases when $m_\sigma$ is below the
limit given by  (\ref{3}), but a quantitative estimate of just how large
this effect is likely to be will require a considerable amount of future
work.

Despite the possibility that the effciency of vorton formation may
have been underestimated by previous work, it still seems unlikely
that vortons can constitute more that a small fraction of the
missing matter in the universe. However this does not mean that 
vortons could not give rise to astrophysically
interesting effects: in particular it has recently been suggested by
Bonazzola and Peter\cite{BP97} that they might account for otherwise
inexplicable cosmic ray events.

The author is grateful to many colleagues, particularly Anne Davis and 
Patrick Peter, for helpful discussions on numerous occasions.

\vfill\eject

\end{document}